\documentclass[prl,twocolumn,amsmath,amssymb,showpacs]{revtex4}
\usepackage{epsfig}
\usepackage{graphicx}

\begin{document}

\title{A matter dominated navigation Universe in accordance with the Type Ia supernova data }

\author{Xin Li$^{1,3}$}
\email{lixin@itp.ac.cn}
\affiliation{${}^1$Institute of Theoretical Physics,
Chinese Academy of Sciences, 100190 Beijing, China}
\author{Zhe Chang$^{2,3}$}
\email{zchang@ihep.ac.cn}
\author{Minghua Li$^{2,3}$}
\email{limh@ihep.ac.cn}
\affiliation{${}^2$Institute of High Energy Physics, Chinese Academy
of Sciences, 100049 Beijing, China\\
${}^3$Theoretical Physics Center for Science Facilities, Chinese Academy of Sciences}

\begin{abstract}
We investigate a matter dominated navigation cosmological model. The influence of a possible drift (wind) in the navigation cosmological model makes the spacetime geometry change from Riemannian to Finslerian. The evolution of the Finslerian Universe is governed by the same gravitational field equation with the familiar Friedmann-Robertson-Walker one. However, the change of space geometry from Riemannian to Finslerian supplies us a new relation between the luminosity distant and redshift. It is shown that the Hubble diagram based on this new relation could account for the observations on distant Type Ia supernovae.
\end{abstract}
\pacs{02.40.-k,98.80.-k}

\maketitle

\section{Introduction}
Einstein's general relativity enables us to come up with a testable theory of the Universe. Hubble\cite{Hubble} first found that the galaxies are receding from us not long after the birth of general relativity. The Hubble's observation indicates that our Universe is expanding. Over the past decade, two groups \cite{Riess,Perlmutter} observing supernovae reported that the luminosity distance can not be explained by a matter dominated Universe. If one accepts the convincible assumption of homogeneity and isotropy of the
Universe which is approximately true on large scales, then the general relativity tells us that we now live in a dark energy dominated Universe and the Universe is accelerated expanding. Dark energy, which has the property of negative pressure, is different from the classical matter and the found particles. A great amount of models have been proposed to study the possible candidates of dark energy and their dynamics \cite{Copeland}. The most famous and acceptable candidate is the cosmological constant. However, the magnitude of cosmological constant $10^{-3}{\rm eV}^4$ is much smaller than the energy density of vacuum in quantum field theory.

The theory of dark energy \cite{Padmanabhan} dominates the modern cosmology in the past decade. This situation is similar to the raise of the theory of ether, which has been considered as direct evidence of the aberration of starlight, an important astronomical effect known since eighteenth century. Although the rapid progress in technology makes the astronomical observations more and more accurate, up to now, there is no direct evidence indicate what the dark energy it is. Since the theory of dark energy is contrived, which requires fine tuning and apparently cannot be tested in the laboratory or solar system, several modified theories of general relativity have been developed as the alternative source of cosmological acceleration \cite{Bludman}.

Einstein first used the Riemann geometry to describe the theory of gravitation. In this Letter, we suppose that the spacetime in large scale may be described by other geometry instead of Riemann geometry. To preserve redeeming feature of general relativity, this geometry must take Riemann geometry as its special case. Fortunately, Paul Finsler proposed a natural generation of Riemann geometry-Finsler geometry.

Finsler geometry as a more general geometry could provide new sight on modern physics. It is of great interest for physicists to investigated the violation of Lorentz symmetry \cite{Kostelecky}. An interesting case of Lorentz violation, which was proposed by Cohen and Glashow\cite{Glashow}, is the model of Very Special Relativity (VSR) characterized by a reduced symmetry SIM(2). In fact, Gibbons, Gomis and Pope\cite{Gibbons} showed that the Finslerian line element $ds=(\eta_{\mu\nu} dx^\mu dx^\nu)^{(1-b)/2}(n_\rho dx^\rho)^b$ is
invariant under the transformations of the group DISIM$_b(2)$. In the framework of Finsler geometry, modified dispersion relation has been discussed\cite{Girelli,Finsler SR}. Also, the model based on Finsler geometry could explain the recent astronomical observations which Einstein's gravity could not. A list includes: the flat rotation curves of spiral galaxies can be deduced naturally without invoking dark matter \cite{Finsler DM}; the anomalous acceleration\cite{Anderson} in solar system observed by Pioneer 10 and 11 spacecrafts should correspond to Finsler-Randers space \cite{Finsler PA}; the secular trend in the astronomical unit\cite{Krasinsky,Standish} and the anomalous secular eccentricity variation of the Moon's orbit\cite{Williams} should be correspond to the effect of the length change of unit circle in Finsler geometry\cite{Finsler AU}.

In this Letter, we present a matter dominated navigation model. The influence of a possible drift (wind) in the navigation cosmolgical model makes the expanding Universe ``accelerated". It is remarkable that the navigation cosmological model is described in the framework of Finsler geometry. We find that the predictions of the navigation cosmological model could account for the observations of Riess and Perlmutter\cite{Riess,Perlmutter} on distant supernovae.

\section{formulism}
Instead of defining an inner product structure over the tangent bundle in Riemann geometry, Finsler geometry is based on
the so called Finsler structure $F$ with the property
$F(x,\lambda y)=\lambda F(x,y)$ for all $\lambda>0$, where $x$ represents position
and $y$ represents velocity. The Finsler metric is given as\cite{Book
by Bao}
 \begin{equation}
 g_{\mu\nu}\equiv\frac{\partial}{\partial
y^\mu}\frac{\partial}{\partial y^\nu}\left(\frac{1}{2}F^2\right).
\end{equation}
In Finsler manifold, there exists a unique linear connection~-~the
Chern connection\cite{Chern}. It is torsion freeness and almost
metric-compatibility,
 \begin{equation}
 \Gamma^{\alpha}_{\mu\nu}=\gamma^{\alpha}_{\mu\nu}-g^{\alpha\lambda}\left(A_{\lambda\mu\beta}\frac{N^\beta_\nu}{F}-A_{\mu\nu\beta}\frac{N^\beta_\lambda}{F}+A_{\nu\lambda\beta}\frac{N^\beta_\mu}{F}\right),
 \end{equation}
 where $\gamma^{\alpha}_{\mu\nu}$ is the formal Christoffel symbols of the
second kind with the same form of Riemannian connection, $N^\mu_\nu$
is defined as
$N^\mu_\nu\equiv\gamma^\mu_{\nu\alpha}y^\alpha-A^\mu_{\nu\lambda}\gamma^\lambda_{\alpha\beta}y^\alpha
y^\beta$
 and $A_{\lambda\mu\nu}\equiv\frac{F}{4}\frac{\partial}{\partial y^\lambda}\frac{\partial}{\partial y^\mu}\frac{\partial}{\partial y^\nu}(F^2)$ is the
 Cartan tensor (regarded as a measurement of deviation from the Riemannian
 Manifold). In terms of Chern connection, the curvature of Finsler space is given as
\begin{equation}
R^{~\lambda}_{\kappa~\mu\nu}=\frac{\delta
\Gamma^\lambda_{\kappa\nu}}{\delta x^\mu}-\frac{\delta
\Gamma^\lambda_{\kappa\mu}}{\delta
x^\nu}+\Gamma^\lambda_{\alpha\mu}\Gamma^\alpha_{\kappa\nu}-\Gamma^\lambda_{\alpha\nu}\Gamma^\alpha_{\kappa\mu},
\end{equation}
where $\frac{\delta}{\delta x^\mu}=\frac{\partial}{\partial x^\mu}-N^\nu_\mu\frac{\partial}{\partial y^\nu}$. The notion of Ricci tensor in Finsler geometry was first introduced by Akbar-Zadeh\cite{Akbar}
\begin{equation}
Ric_{\mu\nu}=\frac{\partial^2\left(\frac{1}{2}F^2 R\right)}{\partial y^\mu\partial y^\nu},
\end{equation}
where $R=\frac{y^\mu}{F}R^{~\kappa}_{\mu~\kappa\nu}\frac{y^\nu}{F}$. And the scalar curvature in Finsler geometry is given as $S=g^{\mu\nu}Ric_{\mu\nu}$.

In standard cosmology, following the cosmological principle, one gets the Friedmann-Robertson-Walker (FRW) metric\cite{Weinberg}. In another word, the spatial part of the Universe is a constant sectional curvature space. Here comes our major assumption, the gravity in large scale should be described in terms of Finsler geometry. In light of the cosmological principle, the Finsler structure of the Universe should be written in such form
\begin{equation}\label{Finsler s}
\bar{F}^2=dt^2-R^2(t)F^2,
\end{equation}
where $R(t)$ is scale factor with cosmic time $t$, the structure $F$ is a constant flag curvature space. By making use of the geometrical terms we mentioned above, one obtains the components of Ricci tensor
\begin{eqnarray}
\label{Ric0}
Ric_{00}&=&-\frac{3\ddot{R}}{R}g_{00},\\
\label{Rici}
Ric_{ij}&=&-\left(\frac{\ddot{R}}{R}+\frac{2\dot{R}^2}{R^2}+\frac{2K}{R^2}\right)g_{ij},
\end{eqnarray}
where the dot denotes a derivative with respect to $t$.

The gravitational field equation in Finsler space should reduce to the Einstein's gravitational field equation while the Finsler space reduce to the Riemannian space. Thus, the symmetrical tensor $G_{\mu\nu}\equiv Ric_{\mu\nu}-\frac{1}{2}g_{\mu\nu}S$ should involve in the gravitational field equation in Finsler space. In general, Finsler space is an anisotropic space. Therefore, it has less Killing vectors than the Riemannian space, and it breaks the Lorentz symmetry\cite{Li}. It means the angular momentum in Finsler space is not a conservative quantity. This fact implies that the energy momentum tensor is not symmetrical in Finsler space. For these reasons, the gravitational field equation in Finsler space should be taken in such form
\begin{equation}
\label{field eq}
G_{\mu\nu}+A_{\mu\nu}=8\pi G(T^s_{\mu\nu}+T^a_{\mu\nu}),
\end{equation}
where $A_{\mu\nu}$ is an asymmetrical tensor and $T^s_{\mu\nu},T^a_{\mu\nu}$ are symmetrical part and asymmetrical part of energy momentum tensor respectively. This general form is agree with the result of Asanov, its gravitational field equation contains the asymmetrical term in Finsler space of Landsberg type\cite{Asanov}. Dealing with field equation in the anisotropic and inhomogeneous cosmology is not a simple staff, and it is hard to find an exact solution of gravitational field equation while its energy momentum tensor involves the non diagonal terms\cite{Misner}. At first glance, we just deal with the symmetrical part of field equation (\ref{field eq})
\begin{equation}
\label{field eqn}
G_{\mu\nu}=8\pi GT_{\mu\nu}.
\end{equation} By making use of the equations (\ref{Ric0}) and (\ref{Rici}), the solution of equation (\ref{field eqn})
is the same with the Einstein's field equation deduced by FRW metric.
Taking the energy-momentum tensor $T_{\mu\nu}$ to be the form of perfect fluid, one can see that the evolution of scale factor $R(t)$ is the same with the Riemannian case. However, since the luminosity distant is related to the spatial geometry of the Universe, one may expect that it is different from the Riemannian luminosity distant.

Unlike Riemann space, a complete classification of the constant flag curvature spaces remain an unsolved problem.  However, by making use of the Zermolo navigation on Riemannian space, Bao {\it et al}. \cite{Bao} gave a complete classification of Finsler-Randers space \cite{Randers} of constant flag curvature. The Zermelo navigation problem \cite{Zermelo} aims to find the paths of shortest travel time in a Riemannian manifold ($M, h$) under the influence of a drift (``wind") represented by a vector field $W$. In standard cosmology, our Universe is very flat now. We may imagine that the Universe is a flat Riemannian manifold with flat Friedmann metric
\begin{equation}
ds^2=dt^2-R_h^2(t)(dx^2+dy^2+dz^2),
\end{equation}
and it influenced by the ``wind" $W$. The relation between the Riemannian manifold ($M, h$) and the Randers metric $F$ is
\begin{equation}
F=\frac{1}{\lambda}\left(\sqrt{\lambda h_{ij}y^iy^j+(W_iy^i)^2}-W_iy^i\right),
\end{equation}
where $W_i=h_{ij}W^j$ and $\lambda=1-h(W,W)$. One should notice that there is a map between the Riemannian space which influence by the ``wind" and the Randers-Finsler space\cite{Gibbons1}. It means the effect of ``wind" already accounted in the gravitational field equation in Finsler space.
The theorem\cite{Bao} of the classification supplies an interesting case where the Randers metric $F$ has constant flag curvature $K$: for $K=-\frac{1}{16}\sigma^2<0$ and $h$ is flat. And $\sigma$ satisfies the constraint $\mathcal{L}_Wh=-\sigma h$, $\mathcal{L}$ denotes Lie differentiation. We set the vector field to be radial $W=\epsilon(t)(x_1\partial x_1+x_2\partial x_2+x_3\partial x_3)$. Then the flag curvature of Randers metric is $K=-\frac{1}{4}\epsilon^2$. After doing coordinate change and taking spherical coordinate, we get the Randers metric as
\begin{equation}
\label{Randers1}
F=\sqrt{\frac{dr^2}{1+\epsilon^2r^2}+r^2(d\theta^2+\sin^2\theta d\phi^2)}-\frac{\epsilon rdr}{1+\epsilon^2r^2}.
\end{equation}
After scale change $r\rightarrow 2r/\epsilon$, the metric $F$ changes as
\begin{equation}
\label{Randers}
F=\sqrt{\frac{dr^2}{1+4r^2}+r^2(d\theta^2+\sin^2\theta d\phi^2)}-\frac{2rdr}{1+4r^2},
\end{equation}
and the cosmic scale factor changes as $R(t)=\frac{2R_h(t)}{\epsilon(t)}$.
Taking the spatial part of the Finsler structure (\ref{Finsler s}) to be of the form (\ref{Randers}), we get metric of the Universe in the framework of Finsler geometry, and the space curvature of the Universe is $-1$. One can deduce directly from such metric that the relation between the redshift $z$ and the scale factor $R(t)$ is the same with the Riemannian case
\begin{equation}
1+z=\frac{R_0}{R(t)}=\frac{1}{a(t)},
\end{equation}
where the subscript zero denotes the quantities given at
the present epoch. Since the non-radial part of the metric (\ref{Randers}) is the same with FRW metric, the luminosity distant in the navigation cosmological model is given as $d_L=R_0r(1+z)$. However, the proper distant is not the case, and it is given as
\begin{eqnarray}
\label{dp0}
d_p&=&\int_0^r\left(\frac{1}{\sqrt{1+4r^2}}-\frac{2r}{1+4r^2}\right)dr\nonumber\\
   &=&\frac{\sinh^{-1}2r}{2}-\frac{1}{4}\ln(1+4r^2).
\end{eqnarray}
The light traveling along the radial direction satisfies the
geodesic equation
\begin{equation}
\bar{F}^2=dt^2-R^2(t)\left(\frac{1}{\sqrt{1+4r^2}}-\frac{2r}{1+4r^2}\right)^2dr^2=0.
\end{equation}
Then, we obtain
\begin{equation}
\label{dp}
d_p=\frac{1}{R_0}\int^1_{(1+z)^{-1}}\frac{da}{a\dot{a}}
\end{equation}
The equations (\ref{field eqn}) and  (\ref{dp}) can be solved. Supposing the Universe is matter dominated (no more cosmological constant and dynamical dark energy), we obtain
\begin{equation}
\label{dp1}
d_p=\log\frac{(\sqrt{1+\Omega_m^{(0)}z}-\sqrt{1-\Omega_m^{(0)}})(1+\sqrt{1-\Omega_m^{(0)}})}{(\sqrt{1+\Omega_m^{(0)}z}+\sqrt{1-\Omega_m^{(0)}})(1-\sqrt{1-\Omega_m^{(0)}})},
\end{equation}
where $H_0$ is the Hubble constant and $\Omega_m$ is the density parameter for matter and satisfies
\begin{equation}
1-\Omega_m^{(0)}=\frac{1}{H_0^2R_0^2}.
\end{equation}
Substituting the equation (\ref{dp1}) into (\ref{dp0}), we obtain the relation between luminosity distant and redshift in the navigation cosmological model
\begin{equation}
H_0d_L=\frac{1+z}{2\sqrt{1-\Omega_m^{(0)}}}\frac{|e^{2d_p}-1|}{e^{d_p}\sqrt{2-e^{2d_p}}}.
\end{equation}

\section{numerical result}
\begin{figure}
\includegraphics[scale=1.0]{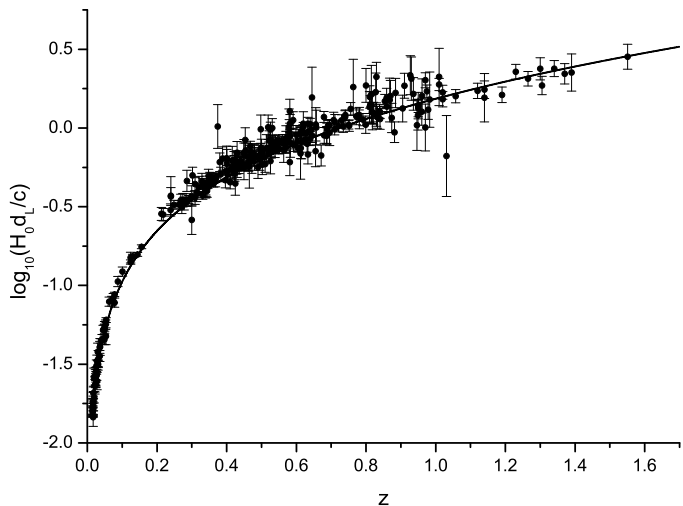}
\caption{The luminosity distant $Log_{10}(H_0d_L/c)$ versus the redshift $z$ for the navigational cosmological model. The data comes from Riess {\it et al}.\cite{Riess} and Hubble Space Telescope (HST)\cite{Kowalski}. And the value of Hubble constant is set as $H_0=67.88km\cdot s^{-1}\cdot Mpc^{-1}$.}
\end{figure}
Here, we present numerical result for the relation between luminosity distant and redshift in the framework of the navigation cosmological model. The best fit curve is shown in Fig.1 with the matter density parameter taken as $\Omega_m^{(0)}=0.92$. And the average value of Chisquare is $1.077588$. Thus, our prediction could account for experiment data given by Riess {\it et al}.\cite{Riess}. It should be noticed that the Hubble constant $H_0$ we took is different from the Hubble constant $H_{h0}$ measured in flat space. The relation for $H$ and $H_h$ is
\begin{equation}
H=H_{h}-\frac{\dot{\epsilon}}{\epsilon}.
\end{equation}
By making use of the value of the Hubble constant measured in flat space $H_{h0}=73\pm0.3~km\cdot s^{-1}\cdot Mpc^{-1}$\cite{Spergel}, we have
\begin{equation}
H_{\epsilon0}\equiv\frac{\dot{\epsilon}_0}{\epsilon_0}=5.42km\cdot s^{-1}\cdot Mpc^{-1}
\end{equation}
This geometrical parameter $H_{\epsilon0}$ represents an ``accelerated" effect provided by the vector field $W$.
\section{conclusion}
Our Letter initiates an exploration of the possibility that the empirical success of the observations of Type Ia supernovae
can be regarded as the influence of a navigational wind. The particles move on Riemnnian manifold and influenced by a vector field
(the ``wind") which proportion to the curvature $-\frac{\epsilon^2}{4}$ of the space of the Universe.
Its geodesic indeed is a Finslerian geodesic. Thus, the observations of Riess {\it et al}.\cite{Riess} on distant supernovae may be
explained by the effect of the ``wind". Our numerical result could account for astronomical observations.
At last, we point out that the age of the Universe is about 9.76Gyr in our model.
This is contradicted with the age ($13.5\pm2Gyr$) of Globular clusters in the MilkyWay\cite{Jimenez}.
In the navigation cosmological model, we only involve the radial ``wind". The non-radial
``wind" should be taken into account in future work, we expect that the effect of the non-radial
``wind" may supply us a reasonable age of the Universe.


\vspace{1cm}
\begin{acknowledgments}
We would like to thank Prof. C. J. Zhu and X. H. Mo for useful discussions. The
work was supported by the NSF of China under Grant No. 10525522 and
10875129.
\end{acknowledgments}

\end{document}